# Nonreciprocal Metasurface with Space-Time Phase Modulation

Xuexue Guo,[1] Yimin Ding,[1] Yao Duan[1], and Xingjie Ni[1,*]

[1]Department of Electrical Engineering, Pennsylvania State University, University Park, PA 16802, USA.

*Correspondence to: xingjie@psu.edu

**Abstract**: Creating materials with time-variant properties is critical for breaking reciprocity that imposes fundamental limitations to wave propagation. However, it is challenging to realize efficient and ultrafast temporal modulation in a photonic system. Here, leveraging both *spatial and temporal phase manipulation* offered by an ultrathin nonlinear metasurface, we experimentally demonstrated nonreciprocal light reflection at wavelengths around 860 nm. The metasurface, with traveling-wave modulation upon nonlinear Kerr building blocks, creates spatial phase gradient and *multi-terahertz temporal phase wobbling*, which leads to unidirectional photonic transitions in both momentum and energy spaces. We observed *completely asymmetric reflections* in forward and backward light propagations within *a sub-wavelength interaction length* of 150 nm. Our approach pointed out a potential means for creating miniaturized and integratable nonreciprocal optical components.

Reciprocity is a fundamental principle rooted in linear physical systems with time-reversal symmetry, requiring that the received-transmitted field ratios are the same when the source and detector are interchanged[1]. However, it is preferable to break reciprocity in many practical applications, such as lasers and full-duplex communication systems, so that back-scattering from defects or boundaries can be avoided[2]. Traditionally, nonreciprocity has been realized through magneto-optic materials which are too bulky and lossy to be integrated into modern photonic systems[3,4]. In addition, nonlinear materials[5-7] are employed to achieve nonreciprocity at a cost of high intensity requirement, but they suffer from poor isolation and are reciprocal to noises[8]. In order to circumvent these limitations, more and more researches have focused on developing materials with time-variant properties in which time-reversal symmetry is explicitly broken to achieve nonreciprocity [1,9]. So far, based on strong electro-optic [10,11], acousto-optic [12-15], or optomechanical effects [16,17] of different materials, proof of concept temporal modulation has been demonstrated at frequencies from kHz to GHz range, which is much lower than the optical frequency as a result of slow carrier injection of electro-optic modulation and low-frequency acoustic or mechanical modes in acousto-optic or optomechanical modulation. In addition, these dynamic systems suffer from limited bandwidth either due to the group velocity mismatch among photonic modes or the intrinsic narrow linewidths of acoustic and mechanical modes. Moreover, they require long interaction lengths to observe the desired effect. Nonreciprocity with *a sub-wavelength interaction length* and an *ultrafast modulation frequency over THz bandwidth* is technically challenging and has not been realized to date.

Here, we experimentally demonstrate a new approach that achieves nonreciprocity on an ultrathin metasurface with space-time phase modulation. A metasurface [18] is an optically thin nanostructured two dimensional material that can manipulate light through its subwavelength-sized building blocks. Designed specifically to achieve a controlled optical response, metasurfaces have been used to create many novel optical devices, including invisibility cloaks [19], flat lenses [20], ultrathin holograms [21], and to explore interesting physical effects [22,23]. While these concepts have opened up a new paradigm for manipulating light with an ultrathin layer, there are fundamental limitations that a spatially modulated metasurface alone cannot overcome. In particular, a time-varying response is required to violate reciprocity in a non-power and non-magnetic-field dependent fashion. The metasurfaces with spatiotemporal modulation in index [24,25] or directly in its phase profile [26] are theoretically proposed recently and are experimentally demonstrated in radio frequencies [27,28], which offers an opportunity to overcome this limitation. We experimentally demonstrate this new dynamic metasurface with an additional fast temporal phase modulation. Our space-time modulated metasurface (Fig. 1a) consists of a set of specially designed nonlinear nanoantennas which not only provides abrupt static phase shifts for the incident light, but also is capable of changing the phase shifts dynamically in the presence of an external travelling wave modulation. In order to introduce the dynamic phase change to the metasurface, we incorporate a heterodyne interference of two coherent waves to generate multi-terahertz (about 2.8 THz) time-varying modulation[29]. This breaks time-reversal symmetry of the metasurface. Our experiments demonstrate nonreciprocal light propagation in free space at around $\lambda = 860$ nm wavelength across *an ultrathin (150 nm) layer*. Furthermore, we achieve *completely asymmetric photonic transitions over a wavelength range around 15 nm (corresponding to a bandwidth of 5.77THz)*. We believe that the space-time modulated metasurface demonstrates a viable way to obtain ultrafast time-variant material responses and potentially motivates a new direction in constructing miniaturized and integratable magnetic-free nonreciprocal optical components.

A conventional spatially modulated metasurface with a phase gradient (e.g. $\varphi = k_s x$) on the surface is capable of imparting additional linear or orbital angular momentum to incident light. This breaks the inversion symmetry and enables full control over the photonic transitions in momentum space. However, this process is linear and time-reversal symmetric, and is inherently reciprocal (Fig. 1, b and c). In contrast, our space-time modulated metasurface has a phase modulation of the form

$$\varphi(x,t) = \Delta\varphi \cos(\Delta\omega t - k_M x) + k_s x, \qquad (1)$$

where $\Delta\varphi$ is the temporal modulation depth, $\Delta\omega$ is the modulation frequency, $k_M$ is the modulation spatial frequency, and $k_s$ is the static phase gradient introduced by the spatial distribution of nanoantennas. This modulation acting upon the incident light field gives an additional space-time varying phase factor expressed as $\exp[i\varphi(x,t)]$. Applying the Jacobi-Anger expansion, this phase term can be rewritten as a series of Bessel functions of the first kind, which enables the reflected field to be expressed as

$$\vec{E}_r(\vec{r},t) = \zeta J_0(\Delta\varphi)\vec{E}_i e^{i(\vec{k}_i \vec{r} + k_s x - \omega_i t)} + i\zeta J_1(\Delta\varphi)\vec{E}_i \left\{ e^{i\left[(\vec{k}_i \vec{r} + k_M x + k_s x) - (\omega_i + \Delta\omega)t\right]} + e^{i\left[(\vec{k}_i \vec{r} - k_M x + k_s x) - (\omega_i - \Delta\omega)t\right]} \right\} \qquad (2)$$

where $\omega_i$ and $\vec{k}_i$ are the frequency and free-space wavevector of the incident wave, $\zeta = \sqrt{\eta}$, and $\eta$ is the static diffraction efficiency of the metasurface. Note that only the zeroth- and first-order Bessel functions are retained since the phase modulation depth $\Delta\varphi$ is small, which leads to negligible contributions from higher-order functions. It is evident from the second term in the right-hand side of equation (2) that a sinusoidal phase component will be decomposed into two photonic transitions (i.e. sidebands) in energy space with resulting frequencies $\omega_r = \omega_i \pm \Delta\omega$, where '+' and '−' denotes an upward and downward transition, respectively. The dynamic phase modulation breaks reciprocity, and leads to time-reversal-asymmetric photonic transitions [30]. Different from the recently reported phonon-mediated nonreciprocal waveguides[14,15] based on indirect interband photonic transitions [30], our system is naturally phase matched for all free space modes and does not require complex design of the acoustic and photonic modes to fulfil stringent momentum/energy matching conditions. Furthermore, our space-time metasurfaces exhibit agile control over both momentum and energy conversions. In order to achieve unidirectional frequency conversion, the metasurface can be designed to either fulfil $k_{ix} + k_s + k_M > k_i$ ($k_{ix}$ is the projection of incident wavevector $k_i$ along the x direction), $2k_M > k_i$, and $-k_i < k_{ix} + k_s - k_M < k_i$ to ensure unidirectional down-conversions (Fig.1 d and e), or fulfill $k_{ix} - k_s - k_M < -k_i$, $2k_M > k_i$ and $-k_i < k_{ix} - k_s + k_M < k_i$ to ensure unidirectional up-conversions. As an example, we depict the case of unidirectional down-conversions in Fig. 1 d and e. The optical paths of forward and backward propagation are shown schematically in the top panel, and the photonic states represented by different color dots are shown in the bottom dispersion diagrams. The photonic transition from the blue dot state to green occurs in the forward propagation, whereas green to red occurs in the backward propagation. With a space-time modulated metasurface, the frequency transitions arise from the parametric processes caused by the temporal modulation, while the momentum transitions arise from both temporal and spatial modulation. As a result, the allowed transitions (i.e. downward transition) can be selected by pushing a given output state (i.e. upward transition) to the forbidden (i.e. non-propagative) region with a unidirectional momentum transfer, $k_s$, provided by the spatial phase modulation of the metasurface. It is worthwhile to note that the reflection angle of the backward propagating light is not necessarily the same as that of the forward propagating light even though they have the same *x* component of the wavevector ($k_x$), because the frequency shift also changes the length of the wavevector. The paths of the incident and returning beam overlap only in the special case of normal incidence (i.e. $k_{ix} = 0$), which is particularly useful for free-space optical isolators. For completeness, we also presented bi-directional photonic transitions on the space-time metasurfaces which also exhibit nonreciprocity (Fig. S2). In all cases, the spatiotemporal phase modulation leads to *completely asymmetric reflections* in forward and backward directions. The back reflected wave does not return to the same state as the forward incident wave, and therefore the process is nonreciprocal.

We used a set of nanobar antennas (Fig. 2a) made of amorphous silicon (α-Si), which has a large Kerr index and low optical loss, as the building blocks of the metasurface. With the adoption of a 50-nm-thick $SiO_2$ spacer layer and a silver back-reflector plate to create a gap resonance, the nanoantenna can induce a large phase shift (over $2\pi$) upon the incident light. The permittivity of the α-Si nanoantennas can be modulated by intense optical field due to the nonlinear Kerr effect. This is an ultrafast process (Fig. S6) and is the key to obtaining the THz temporal phase modulation. Subsequently, the light-induced permittivity change of the nanoantennas will detune their resonances and therefore lead to a change in the phase shift upon the incident light at the operational wavelength. The static phase shifts ($\varphi$) and the changes in the phase shifts ($\Delta\varphi$) induced by the pump light ($\lambda_{pump}$ = 800 nm, $I_{pump}$ = 15 GW/cm$^2$) are simulated separately and mapped out for the selection of designs in a two-dimensional parameter space spanned by nanoantenna dimensions $l_x$ and $l_y$. Three different nanoantennas having static phase shifts covering $2\pi$ range while preserving a uniform phase shift change under pump light illumination were chosen to build a supercell of the metasurface. The parameters for the nanoantenna designs used in our experiment are indicated as red diamonds in Fig. 2b. The spatial distribution of these nanoantennas create a static phase gradient $k_s = 2\pi/p_x$, where $p_x$ is the supercell period in the $x$ direction. Note that *the nanoantennas transform the small permittivity modulation from the Kerr effect into relatively large dynamic phase modulation*, leading to efficient nonreciprocal photonic transition at moderate pump power. Moreover, the local field is enhanced by those resonant elements. As a result, the designed nanoantennas significantly boost the temporal phase modulation depth.

We fabricated the metasurface as shown in Fig. 2c and characterized its linear performance with a $k$-space imaging microscope (Fig. S5a). Our metasurface attains an overall diffraction efficiency of 84% *experimentally* around the designed operational wavelength $\lambda$ = 860 nm (Fig. S5c). The measured reflection angles agree with the theory at different wavelengths and different supercell period (Fig. 2d).

Next, we developed a fast temporal modulation technique which uses a heterodyne interference between two laser lines that are closely spaced in frequency (Fig. 3) [31]. In contrast to a homodyne interference setup, the heterodyne interference pattern results in a traveling wave intensity distribution given by $I(x,t) = I_0[1 + \cos(\Delta\omega t - k_M x)]$, where $k_M = 2\pi/\Lambda_M$ ($\Lambda_M$ is the period of the interference fringes) and $\Delta\omega = \omega_{p2} - \omega_{p1}$ is the frequency difference between the two pump laser lines. Projecting this interference pattern on the designed nonlinear metasurface imprints a travelling wave phase profile, $\Delta\varphi \cos(\Delta\omega t - k_M x)$, onto the reflected wave. Together with the spatial phase modulation $\varphi = k_s x$ created by the distribution of nanoantennas on the metasurface, we obtained a space-time modulation of the form described by equation (1). Therefore, the incident wave acquires the energy and momentum shifts when reflected by the metasurface. This process shares some similarities with conventional four-wave-mixing in a homogeneous nonlinear material [32]. However, there are subtle but essential differences between them. In our case, we used the meta-atoms for tailoring locally both the linear and nonlinear responses – collectively the metasurface exhibits large temporal phase wobbling under a driving field in addition to a static phase shift, which cannot be achieved using a natural material.

The experimental setup for creating the temporal modulation is illustrated in Fig. 3b and described in the Methods session. In our experiment, the center wavelength difference was around 6 nm and the frequency difference was around 2.8 THz (Fig. 3c). In addition, $k_M$ was adjusted by changing the angle between the two pump beams impinging onto the metasurfaces. Fourier transform analysis of the interference pattern shows a $k_M$ equal to $0.54k_{probe}$ (Fig. 3d), where $k_{probe}$ is the length of the free-space wavevector of the probe beam.

To demonstrate the nonreciprocal light propagation, the metasurface with $k_s = 0.72k_{probe}$ was imprinted by the interference pattern at a peak pump intensity around 1 GW/cm$^2$, which generated a dynamic phase modulation with $\Delta f = 2.8$ THz and $k_M = 0.54k_{probe}$. For the forward propagation experiment, the 860 nm probe light hit the metasurface at normal incidence and resulted in a reflected wave with a shift in the energy-momentum space. This shift was captured by collecting the reflected spectra from a fiber aperture scanning spatially on the Fourier plane (along the $k_x$ direction) of the focusing lens before the metasurface. Fig. 4a displays a static diffraction at 348.6 THz and $0.72k_{probe}$ determined by $k_s$, and a down-shifted signal at 345.8 THz and $0.18k_{probe}$ produced by the spatiotemporal modulation induced $\Delta f$, $k_M$, and $k_s$. The 0$^{th}$-order reflection can also be detected, as the diffraction efficiency dropped due to edge effects and slight polarization misalignment. The aspheric lens used to focus the pump and probe beams and to collect converted signals has an effective NA of 0.76; $k_x/k_{probe}$ is therefore bounded by the limited collection angle. On the other hand, the accumulated $k_x/k_{probe} \sim 1.26$ of the upward transition is greater than unity. Therefore, it is evanescent and cannot carry energy away from the metasurface. For the backward propagation experiment, we sent in a probe beam ($f = 345.8$ THz and $k_x = -0.18k_{probe}$) with the same frequency as the previous down-converted signal but the opposite direction onto the metasurface. We observed another downward transition at 343.0 THz exit along the normal direction (Fig. 4b). Similarly, the upward transition is nonexistent since its accumulative $k_x$ is greater than $k_{probe}$. Therefore, the backward propagating light cannot return to the initial state. We also did a fine scan over the $\omega$-$k_x$ regions where high order conversions may exist. But no converted signals were observed since these processes have much lower efficiency. In addition, we did a control experiment on amorphous silicon film (thickness ~ 150nm) using similar pump intensity, but no converted signal was detected. We experimentally realized nonreciprocal light reflections on the space-time phase modulated metasurface. The experimental results are also confirmed by our finite-difference time domain (FDTD) simulations (Fig. S3). In addition, we investigated the scattering matrix of our space-time phase modulated metasurface. It is asymmetric and hence is a direct evidence that our system is nonreciprocal (see S4 of supplementary information). In order to experimentally show the nonreciprocal operation wavelength range of our metasurface, we conducted additional measurements on the nonreciprocal processes by mapping out the dispersion diagrams (Fig. S11) with a narrowband probe of which the center wavelength is ranging from 854 to 914 nm for both forward and backward reflections. By extracting the conversion efficiencies at the tested wavelengths (Fig. S12), a 3-dB bandwidth (full-width at half-maximum, FWHM) of approximately 5.77 THz was demonstrated (see S6 of supplementary information). It is worth noting that in contrast to the waveguide-based systems, the bandwidth of our space-time metasurface is not constrained by the phase-matching conditions. Our experimentally obtained bandwidth is at least one order of magnitude greater than the largest ones reported (a few hundred GHz)[10,15] on time-dependent nonreciprocal systems.

Using our space-time metasurface, we are able to control *independently* the static phase gradient $k_s$, the dynamic spatial frequency $k_M$, and the temporal modulation frequency $\Delta\omega$, which provides unpreceded flexibility in manipulating the photonic transitions. As shown in Fig. 5, by changing $k_s$ we can selectively enable downward or upward transitions. Additionally, we demonstrated nonreciprocal reflections with arbitrary transverse momenta by tuning both $k_s$ and $k_M$ along the metasurfaces (Fig. S10). Moreover, we showed that the modulation frequency can be changed by adjusting the frequency splitting between the two pumps (Fig. S9a). It is worthy to note that this level of controllability has not been achieved in previous time-variant nonreciprocal systems[10-17,33].

We experimentally demonstrated nonreciprocal light reflection based on the space-time modulated nonlinear metasurface. The heterodyne interference created by frequency-shifted pump beams provides robust and controllable spatiotemporal modulation, of which $\Delta\omega$ and $k_M$ can be readily tuned as desired. The spatiotemporal phase modulation greatly expands the functions of conventional static phase gradient metasurfaces, providing an additional degree of freedom for manipulating the temporal properties of light and achieving nonreciprocal light propagation. Particularly, we achieved in our experiment *2.8 THz modulation frequency*, a huge step towards optical frequencies, and about *5.77 THz 3-dB bandwidth* which is orders of magnitude greater than current time-variant nonreciprocal systems to the best of our knowledge.

Although travelling wave modulation induced nonreciprocity has been already explored in theory and demonstrated in waveguide systems, such as optical fiber and silicon waveguides, experimental realizations in free space optical systems remain scarce due to lack of fast efficient dynamic modulation technique. Previous demonstrations rely on long interaction lengths (centimeter to meter scale) to amplify the weak dynamic modulation, which make it difficult to be realized in a free space optical system. However, through the use of nonlinear optical metasurface to greatly enhance the dynamic modulation strength, the optical nonreciprocity is achieved at a subwavelength-scale interaction length on the metasurfaces, making this approach potentially compatible with integrated nanophotonic and quantum optical systems. Moreover, by using dynamic phase modulation of metasurface working in free space, our operation bandwidth is not limited by the group velocity mismatch among guided modes[10,15]. Instead, it is the resonant linewidth of the metasurface that determines the nonreciprocal bandwidth, which is approximately in the tens of THz range.

The conversion efficiency, which is defined as $P_{signal}/P_{probe}$ and is proportional to $J_1(\Delta\varphi)^2$, is on the order of $2.5 \times 10^{-4}$ in our experiments due to limited experimental conditions. The temporal mismatch (Fig. S6 a) between the pump (~ 400 fs) and probe (~ 2 ps) beams limits the effective interaction time between probe and the dynamic modulation. Due to bad spatial profile of the probe beam and lens aberration, only about 20% of probe beam overlaps with the pump beams, resulting in reduced effective interaction area. In addition, due to the high repetition rate of our laser system (80-MHz), we need to keep a low pump intensity to avoid significant thermal effect (Fig. S14).

The conversion efficiency of our space-time metasurface can be improved in the following aspects. We find that the conversion efficiency increases super-linearly with the pump intensity in our system (Fig. S14). With pump light of low repetition rate and optimized spatiotemporal overlap, we will be able to achieve higher conversion efficiency. In addition, highly nonlinear materials, e.g. indium tin oxide (ITO) which has gigantic third-order nonlinearity in its epsilon-near-zero region [34,35], can be used to greatly increase the dynamic phase modulation depth. Furthermore, optimized nanoantennas design, such as doubly resonant nanoantennas, will improve the temporal phase modulation depth and relax the pump power requirement. Besides, by stacking a pair of space-time phase modulated metasurfaces to form a cavity or integrating metasurface with resonator structures, *i.e.* micro-ring resonators, the effective interaction length/time will be tremendously increased without sacrificing the device footprint, leading to enhanced nonreciprocal mode conversion efficiency. (Detailed discussions of the aforementioned methods are included in S7 of supplementary information.)

In conclusion, we have demonstrated nonreciprocal light reflection on an ultrafast space-time phase modulated metasurface. This approach exhibits excellent flexibility in controlling optical mode conversions. Moreover, the dynamic system features broken time reversal symmetry compatible with topological insulators, which inspires a new route for isolation using topologically protected edge states [36] that are immune to disorders. We believe it will provide a new platform for exploring interesting physics in time dependent material properties and will open a new paradigm in the development of scalable and integratable nonreciprocal devices.

## Methods

**Sample fabrication.** A 200-nm layer of silver was deposited onto a silicon substrate with a 5-nm Ge adhesion layer by electron-beam (e-beam) physical vapor deposition (SEMICORE E-Gun Thermal Evaporator). A 50-nm SiO2 dielectric spacer layer and 150-nm amorphous silicon layer were then grown by plasma enhanced chemical vapor deposition (PECVD). The metasurface nanoantennas were created using a sequential process of e-beam lithography (EBL), lift-off of chromium mask and inductively coupled plasma - reactive ion etching (ICP-RIE). In the EBL process, we employed a 1:1 diluted ZEP 520A e-beam resist to achieve high resolution. The total pattern size was 200 by 200 $\mu m^2$, written using a Vistec 5200 100 kV ebeam writer. A chlorine-based plasma RIE recipe involving $Cl_2$ and Ar gas was used to etch amorphous Si, creating the nanoantennas. The sample was finally immersed in a chrome etchant to remove the chrome mask.

**Simulation method.** We developed a two-step full-wave model in a commercial finite element method solver (COMSOL Multiphysics) to obtain the nonlinear Kerr-effect-induced phase shift change ($\Delta\varphi$). The first step iteratively calculates pump-beam-modified effective permittivity of the structure. The second step simulates a weak probe beam incident onto the structure with the calculated effective permittivity in the first step. By varying the geometrical parameters of the nanoantennas, we obtained the corresponding phase shift changes.

We also performed time-domain simulations using finite-difference time-domain (FDTD) to study the spatiotemporal phase modulation induced nonreciprocal photonic transitions. We updated the fields in each time step taking into account the temporal changes in permittivity. By varying the spatiotemporal modulation parameters $\Delta\varphi$, $k_s$, $k_m$ and $\Delta\omega$, we simulated different conditions corresponding to the experimental demonstrations.

Please refer to session S1 of the supplementary infomation for more detailed information on the simulation method.

**Nonreciprocal reflection experiment.** To perform measurements of nonreciprocal reflection, the experiment setup is shown in Fig. 3b. In order to create two pump beams with a frequency separation for the travelling-wave modulation, the output of a Ti:Sapphire pulsed laser radiation (140-fs pulse width, 80-MHz repetition rate) at $\lambda = 800$ nm was dispersed by a transmission grating and then spatially separated by a customized split mirror with a variable-size block attached in the center to tune the frequency separation. The spatially separated beams were then reflected back through the grating and recombined into two pump beams with shifted central frequencies. By adjusting the delay line 1, the temporal delay between the two pump beams can be tuned. To create the probe beam, the other small portion of the Ti:Sapphire pulsed laser radiation was sent to a nonlinear photonic crystal fiber (PCF) to generate a supercontinuum. The probe light at 860 nm wavelength was selected using a monochromator. By adjusting the delay line 2, the probe can be synchronized with the pumps. An aspheric lens with effective NA of 0.76 focused the three beams onto the metasurface that was mounted on a three-dimensional (3D) translation stage. Due to aberration of the aspheric lens, the focal spot of the pump beams is an ellipse with major and minor axis lengths of 50 μm and 45 μm, respectively. The reflected signal was directed to a fiber coupler by a D-shaped pickup mirror, of which the position was adjusted by a linear translation stage to collect the *k*-space information of the output signal.

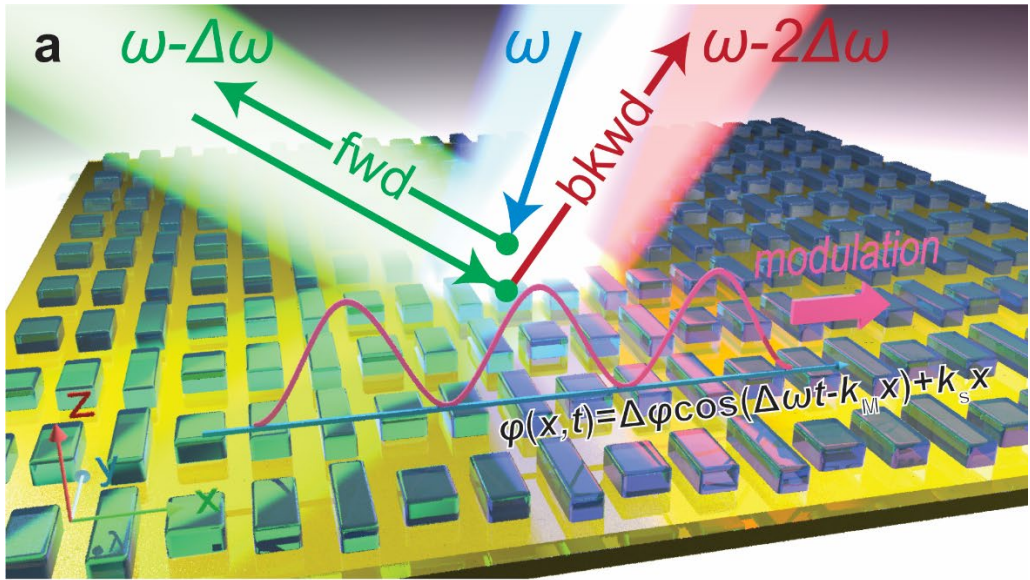
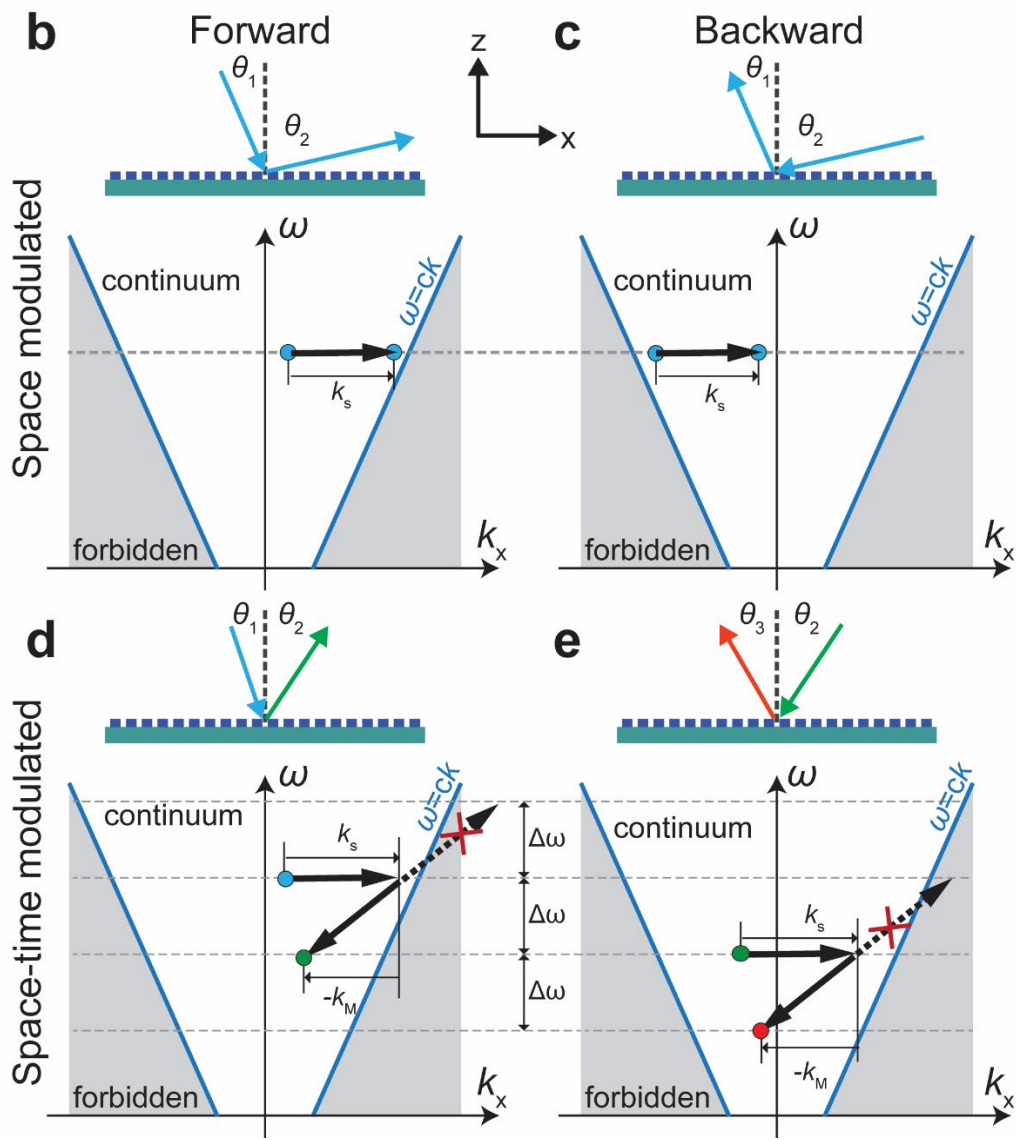

**Fig. 1. Working principle of a nonreciprocal space-time phase modulated metasurface. a**, An illustration showing the concept of a space-time phase modulated metasurface consisting of resonating dielectric nanoantennas operating in the reflection mode. A travelling phase modulation in a sinusoidal form is superposed on the designed phase gradient along the *x* direction. Light impinging on the metasurface with frequency $\omega$ is converted to a reflecting beam with frequency $\omega - \Delta\omega$ due to the parametric process arisen from dynamic phase modulation, while the back-propagating beam with frequency $\omega - \Delta\omega$ will be converted to $\omega - 2\Delta\omega$ instead of $\omega$, resulting in a nonreciprocal effect. **b**, **c** and **d**, **e**, Comparison between a regular space modulated metasurface (b) and (c) and a space-time phase modulated metasurface (d) and (e). A regular space modulated metasurface only supports symmetric forward (b) and backward (c) reflections as shown in the dispersion diagrams. There is no frequency conversion (i.e. no transition in energy space). The process is reciprocal and the forward and backward beams share the same trajectory. In contrast, a space-time phase modulated metasurface supports asymmetric forward (b) and backward (c) reflections. It not only offers additional momentum along x direction to the reflected light but also converts its frequency. In either forward or backward case, the upward transition is forbidden because the resulting wavevector will be too large to be supported in free space, resulting in unidirectional photonic transitions in both energy and momentum spaces. Therefore, the trajectories of beams differ and reveal nonreciprocity effect. $k_s$ is the linear momentum introduced by the spatial phase modulation and $k_M$ is the additional linear momentum introduced by the temporal phase modulation.

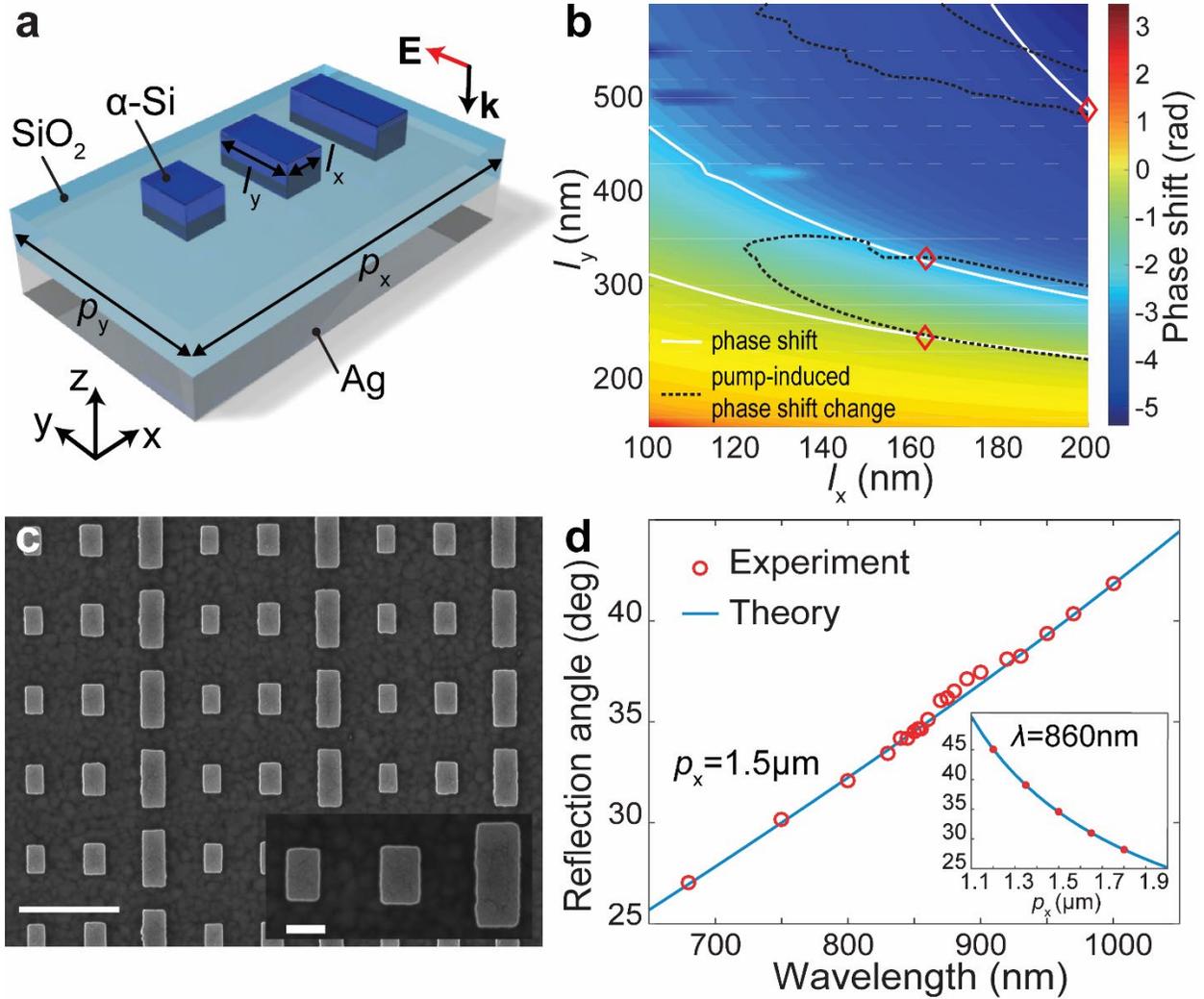

**Fig. 2. Design and characterization of the metasurface. a**, A 3D illustration of a unit cell of the metasurface which consists of three α-Si nanobar antennas. The thickness of the sliver ground plate, the SiO₂ spacer layer, and the α-Si nanoantennas are 200 nm, 50 nm, and 150 nm, respectively. **b**, Calculated phase shifts (surface plot) of reflected light in a 2D parameter space spanned by $l_x$ and $l_y$. It is overlaid by contour lines showing the pump-induced phase shift change of 0.32 radians (black dashed line) when illuminated with pump light at intensity 15 GW/cm². The white lines are the contours indicating the evenly spaced phase shifts in the static condition. Three different nanoantennas which cover $2\pi$ static phase shifts with an interval of $2\pi/3$ are chosen as the building blocks to construct the metasurface, as marked by the red diamonds intersecting the two types of contour lines. **c**, Field emission scanning electron microscopy (FESEM) image of a fabricated α-Si metasurface. Scale bars in the main figure and the inset are 1μm and 200nm, respectively. **d**, Measured (red circles) and calculated (blue line) anomalous reflection angles of the metasurface at wavelengths ranging from $\lambda$ = 680 nm to 1000 nm at normal incidence, with $p_x$ = 1.5 μm. The inset shows the anomalous reflection angles for samples with $p_x$ varying from 1.2 μm to 1.8 μm at $\lambda$ = 860 nm. The experimental results agree well with the theory.

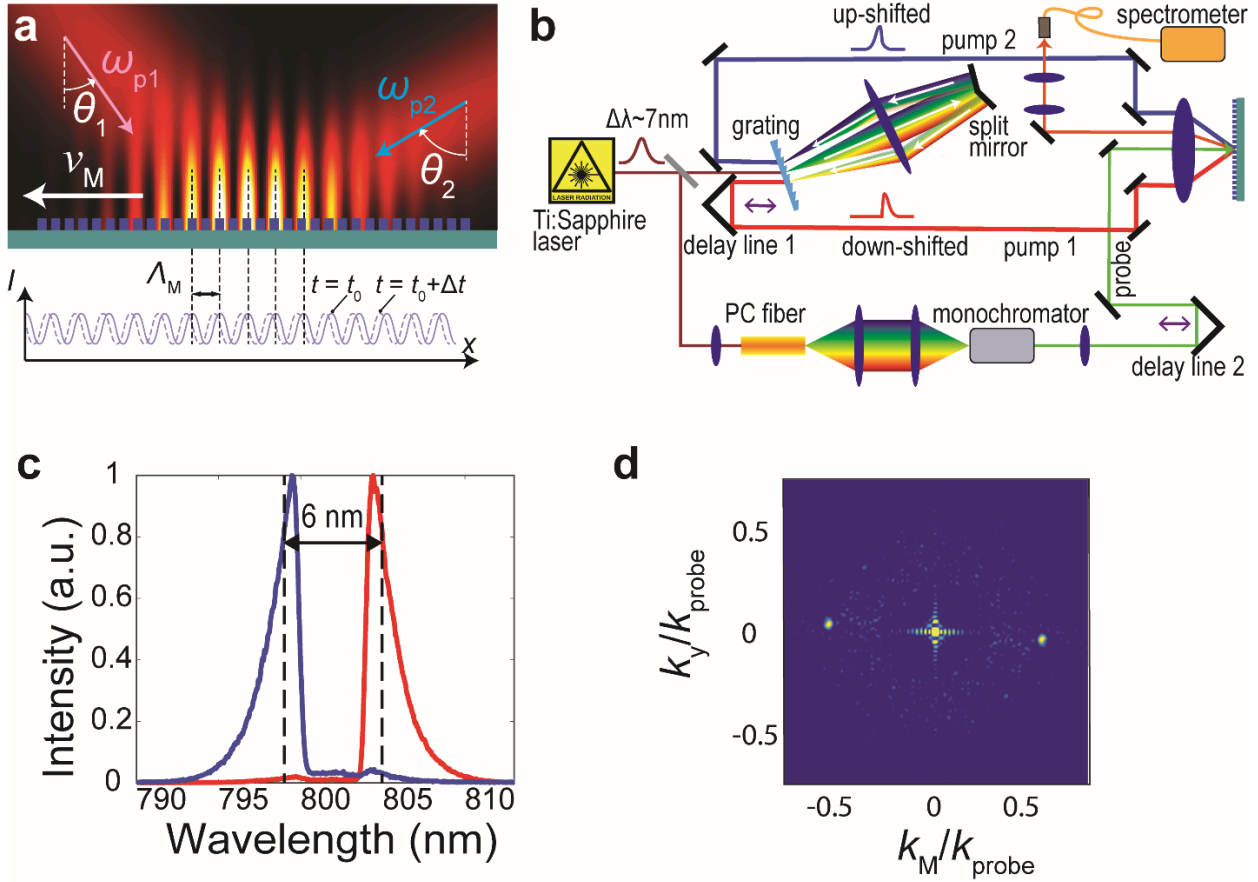

**Fig. 3. Experimental demonstration of controllable space-time phase modulation. a**, A schematic of two pump beams with closely spaced frequencies $\omega_{p1}$ and $\omega_{p2}$ impinging on the metasurface with incident angles $\theta_1$ and $-\theta_2$, respectively. They produce travelling interference fringes with a period of $\Lambda_M$ and a speed of $v_M = \Delta\omega / k_M$ on the metasurface. The nanoantennas on the metasurface exhibit a time-variant change in the phase shifts induced by the travelling interference fringes. **b**, A schematic of the experimental setup. The output of a Ti:Sapphire femtosecond pulsed laser at 800 nm is split into two beams: one is directed through a transmission grating to generate frequency-shifted pump beams; the other is sent to a photonic crystal fiber (PCF) to create a wavelength-tunable probe beam. Two delay lines are employed in order to achieve the temporal synchronization among the three beams. An aspheric lens focuses pump and probe beams onto the metasurface. The reflected signal is picked up by a D-shaped mirror and detected by a spectrometer. We map out the frequency and momentum of the reflected signal by monitoring the collected spectra across the Fourier plane of the aspheric lens. **c**, Spectra of the two pump beams, showing a wavelength difference of 6 nm (corresponding to $\Delta f$ of 2.8 THz). **d**, The 2D Fourier transform of the interference pattern of the two pump beams, revealing a $k_M$ equal to $0.54k_{probe}$, where $k_{probe}$ is the free space wavevector of the probe light at $\lambda = 860$ nm.

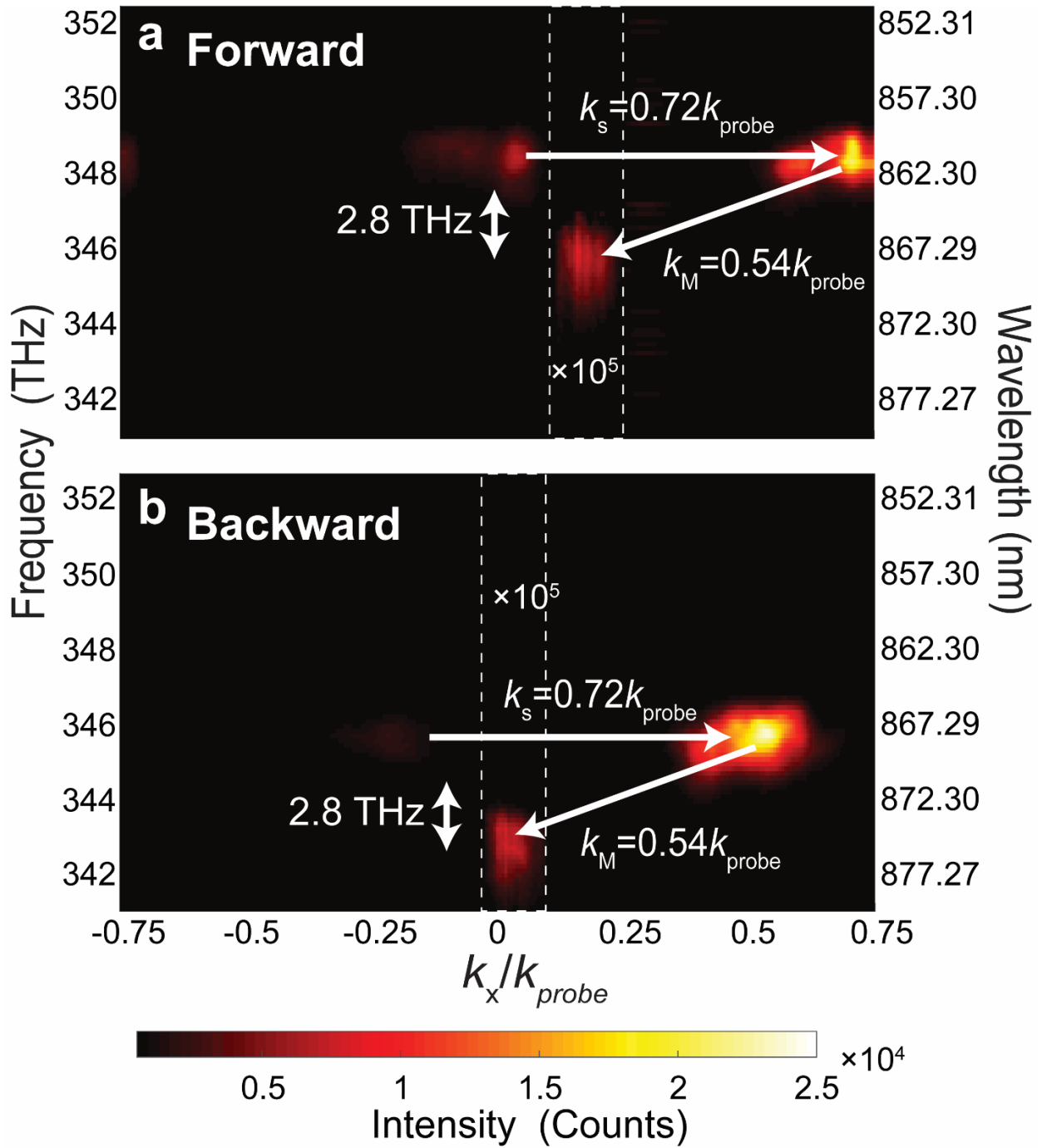

**Fig. 4. Experimental demonstration of nonreciprocal light reflection on the space-time phase modulated metasurface. a**, Under dynamic modulation with $\Delta f = 2.8$ THz and $k_M = 0.54 k_{probe}$, the energy-momentum diagram of the normal-incident probe beam ($f = 348.6$ THz) on the metasurface ($k_s = 0.72 k_{probe}$) shows a downward converted signal at $f = 345.8$ THz and $k_x =$

$0.18k_{probe}$. **b**, In the backward case, a probe beam having the same frequency with previous signal ($f$ = 345.8 THz) but the flipped direction ($k_x = -0.18k_{probe}$) was sent onto the metasurface. The energy-momentum diagram shows a further down-shifted signal at $f$ = 342.0 THz exiting in the normal direction ($k_x = 0$). In both cases, the converted signals are magnified by $10^5$ for better illustration (the magnified regions are enclosed by the dashed white boxes). These results are perfectly matched with our theoretical prediction depicted in Fig. 1d and e. In the regions of interest (where all the possible photonic transitions exist), finer scanning steps and long spectrometer integration time were used to ensure to detect converted signals. In the regions with possible upper sidebands or higher-momentum sidebands, no signal was detected even with magnification. Therefore, here only the regions where the 1$^{st}$-order conversions occur are magnified.

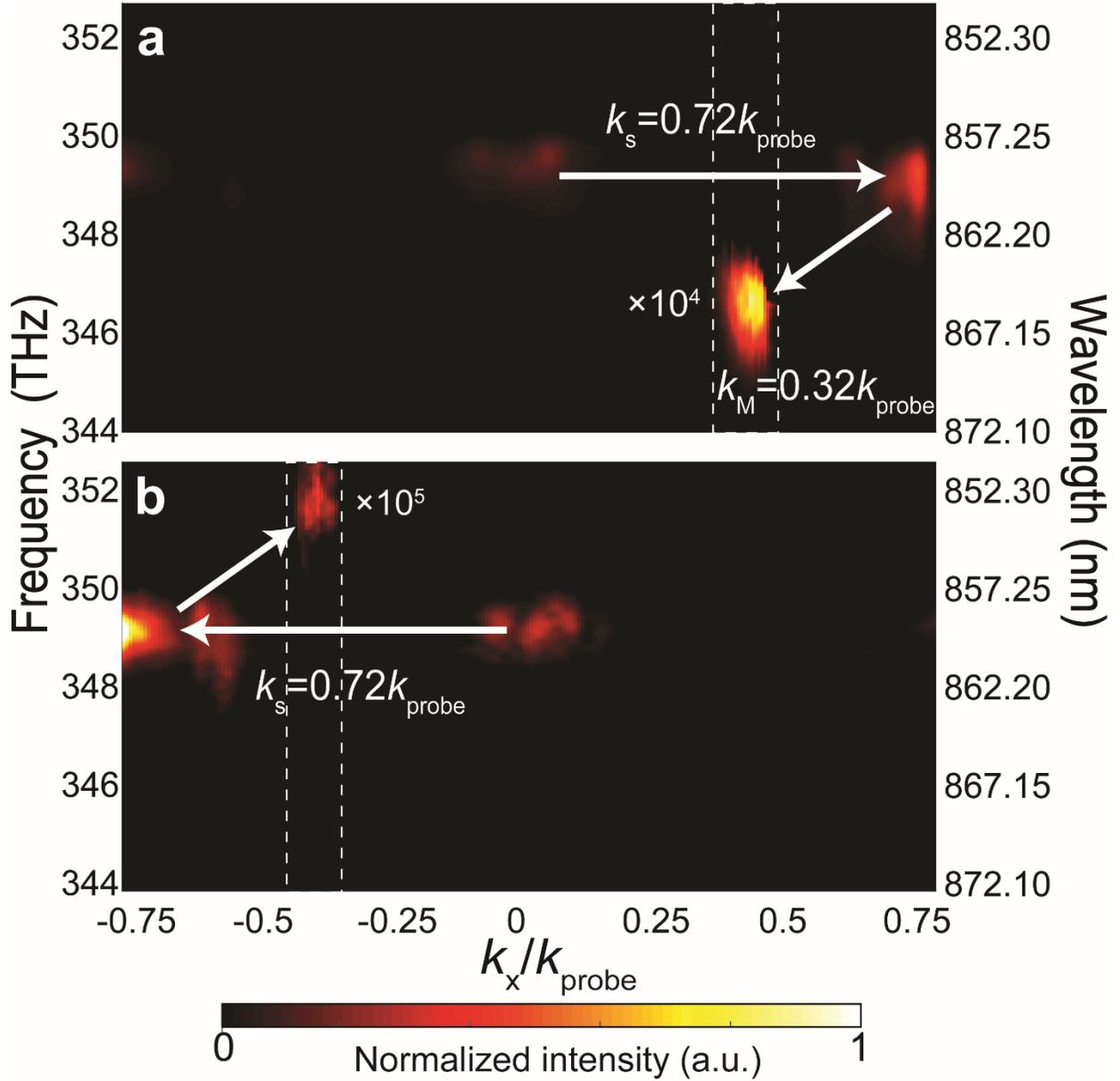

**Fig. 5. Experimental demonstration of direction selectivity of the photonic transitions. a**, Only downward photonic transition occurs on a metasurface ($k_s = 0.72k_{probe}$) modulated by $k_M = 0.32k_{probe}$ and $\Delta f = 2.8$ THz. The converted signal is magnified by $10^4$ for better illustration. **b**, On the contrary, with the same temporal phase modulation on a metasurface with $k_s = -0.72k_{probe}$, only upward photonic transition takes place. The converted signal is magnified by $10^5$ for better illustration (the magnified regions are enclosed by the dashed white boxes).